\documentclass[twocolumn,pra,aps,showpacs]{revtex4}

\usepackage{psfrag,graphicx}

\usepackage{ae,aecompl}

\usepackage{overpic}

\usepackage{epsfig}

\usepackage{dcolumn}

\usepackage{amsmath,amssymb}

\usepackage{bm}

\usepackage[dvips]{color}

\definecolor{mygrey}{gray}{0.35}

\definecolor{mygreen}{rgb}{0.85,1,0.9}

\definecolor{myzard}{cmyk}{0,0,0.05,0}

\definecolor{mywhite}{rgb}{1,1,1}

\definecolor{myred}{rgb}{1,0,0}

\def\dd{\mathord{\rm d}}

 \def\ii{\mathord{\rm i}}

\def\bra#1{\langle#1|} \def\ket#1{|#1\rangle}

\begin{document}

\title[Short Title]{A Lorentz Invariant Pairing Mechanism: Relativistic Cooper Pairs}

\author{A. Bermudez$^1$ and M.A. Martin-Delgado$^1$ }

\affiliation{ $^1$Departamento de F\'{\i}sica Te\'orica I,
Universidad Complutense, 28040. Madrid, Spain.}

\begin{abstract}

We study a Lorentz invariant pairing mechanism that arises when
two relativistic spin-1/2 fermions are subjected to a Dirac string
coupling. In the weak coupling regime, we find remarkable
analogies between this relativistic bound system and the well
known superconducting Cooper pair. As the coupling strength is
raised, quenched phonons become unfrozen and dynamically
contribute to the gluing mechanism, which translates into novel
features of this relativistic superconducting pair.

\end{abstract}


\maketitle

\section{Introduction}
\label{sectionI}

A large class of superconducting materials can be accurately
described by the BCS theory~\cite{BCS}, which is based upon two
major contributions. First, Fr\"{o}lich showed how the coupling
between electrons and crystal phonons leads to an effective
attractive interaction between the electrons~\cite{frohlich}.
Inspired by this result, Cooper discussed how any attractive
interaction can bind a couple of electrons which lay around a
filled Fermi sea~\cite{Cooper}. This bound system, known as a
Cooper pair, is responsible for several intriguing properties
displayed by superconductors, which can be described as a
many-body coherent state where electrons above the Fermi surface
are bounded in pairs.

\begin{figure}[!hbp]
\centering
\begin{tabular}{cc}
\epsfig{file=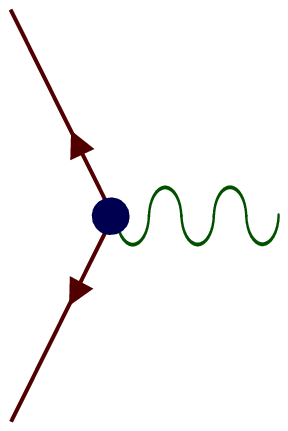,width=0.3\linewidth,clip=}&
\hspace{5ex}\epsfig{file=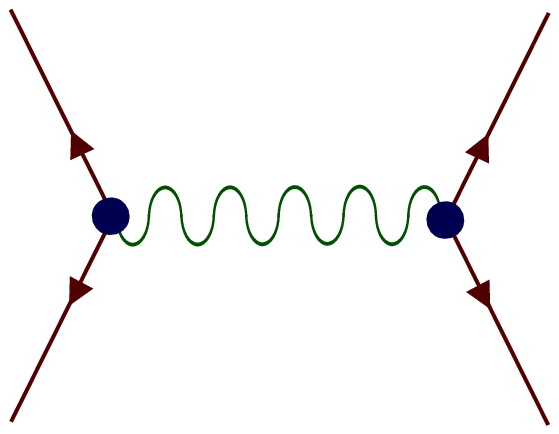,width=0.6\linewidth,clip=} \\
\end{tabular}
\caption{(left) Electron-phonon interaction in the language of
Feynman diagrams. (right) Effective electron-electron attractive
interaction due to the exchange of a lattice phonon.}
\label{feynman_diagram}
\end{figure}

The oversimplified picture developed by Cooper captures the
essence of the underlying physical phenomena occurring in
superconducting solids. In the same spirit, we study a simple
model of two relativistic fermions with an effective attractive
interaction. In order to maintain the similarities with the Cooper
problem, we must fulfill the two following requirements:

\noindent \textbf{Phonon gluing mechanism:}  In a
relativistic scenario, the simplest phonon-like coupling is
modeled by a Dirac string coupling mechanism where the vibrations
of the string describe the lattice phonons (see
fig.~\ref{feynman_diagram} left). This interaction is introduced
by a non-minimal coupling procedure in the free Dirac equation
\begin{equation}
\label{1_body_dirac_oscillator} \ii \hbar \frac{\partial
\ket{\Psi}}{\partial t}=\left(c\boldsymbol{\alpha}(\textbf{p}-\ii
m\omega\beta\textbf{r})+mc^2\beta\right)\ket{\Psi},
\end{equation}
where $\ket{\Psi}$ stands for the Dirac 4-component spinor,
$\textbf{p}$ represents the momentum operator, and $c$ the speed
of light. Here
$\beta=\text{diag}(\mathbb{I}_2,-\mathbb{I}_2),\alpha_j=\text{off-diag}(\sigma_j,\sigma_j)$
are the Dirac matrices in the standard representation with
$\sigma_j$ as the usual Pauli matrices \cite{greiner_book}. This
Dirac string coupling $\textbf{p}\to\textbf{p}-\ii
m\omega\beta\textbf{r}$ was introduced in~\cite{imc67,moshinsky}
as a relativistic extension of the harmonic oscillator, usually
coined as the Dirac oscillator, where $\omega$ represents the
oscillator's frequency. In our picture, this frequency effectively
describes the lattice vibrations and its coupling to the fermionic
degrees of freedom.

\noindent \textbf{Two-fermion binding:} Regardless of the coupling
strength, we shall show that such an effective string coupling
binds relativistic fermions in pairs (see
fig.~\ref{feynman_diagram} right). A Lorentz invariant extension
of the Dirac string Hamiltonian in
Eq.\eqref{1_body_dirac_oscillator} to two-fermion systems is
possible
\cite{barut,moshinsky_2Body,moshinsky_meson,moshinsky_pp+}, which
in the center of mass reference frame reads as follows
\begin{equation}
\label{2body_DO} H_{\text{3D}}=\frac
{c}{\sqrt{2}}(\boldsymbol{\alpha}_1-\boldsymbol{\alpha}_2)(\textbf{p}-\ii
m \omega\beta_{12}\textbf{r})+mc^2(\beta_1+\beta_2),
\end{equation}
where
$\boldsymbol{\alpha}_1=\boldsymbol{\alpha}\otimes\mathbb{I}_4$,
$\boldsymbol{\alpha}_2=\mathbb{I}_4\otimes\boldsymbol{\alpha}$,
$\beta_1=\beta\otimes\mathbb{I}_4$,
$\beta_2=\mathbb{I}_4\otimes\beta$, and
$\beta_{12}=\beta\otimes\beta$ represent the generalization of the
Dirac matrices in the two-body Hilbert space. Here
$\textbf{p}:=(\textbf{p}_1-\textbf{p}_2)/\sqrt{2}$, and
$\textbf{r}:=(\textbf{r}_1-\textbf{r}_2)/\sqrt{2}$ stand for the
relative momentum and position operators.

In this work, we study the binding properties of this two-body
relativistic Hamiltonian, and discuss under which circumstances an
analogy to Cooper pairs can be performed. Phonons in this
relativistic system are dynamical and always provide a pairing
mechanism, as we shall see. Thus, there is no need to invoke a
many-body effect through the Pauli principle as in the original
Cooper pair scenario. In fact, there are real materials which
deviate from standard BCS theory. In BCS, phonons are quenched and
their effect appears as a pairing energy scale, but they are not
explicit in the Hamiltonian. There is an extension of the BCS
theory that accounts for the effects of dynamical phonons, known
as the Migdal-Eliashberg theory \cite{migdal,eliashberg,carbotte}.
Our relativistic fermionic pairing mechanism is thus closer to
this latter treatment.

We shall restrict ourselves to a two-dimensional system, where an
exact solution is derived and several interesting properties can
be neatly discussed. Two spatial dimensions are also natural for
other superconducting materials like the cuprates
\cite{bednorz_muller}. In this case, the Dirac matrices reduce to
the usual Pauli matrices
$\alpha_x=\sigma_x,\alpha_y=\sigma_y,\beta=\sigma_z$, and the
relativistic 1-body state $|\Psi\rangle$ can be described by a
2-component spinor. The 2-body relativistic Hamiltonian in two
dimensions can be written as follows
\begin{equation}
\label{2D_2body_DO}
\begin{split}
H_{\text{2D}}= & \frac
{c}{\sqrt{2}}(\sigma_x\otimes\mathbb{I}_2-\mathbb{I}_2\otimes\sigma_x)(p_x-\ii
m \omega\sigma_z\otimes\sigma_zx)\\
+&\frac
{c}{\sqrt{2}}(\sigma_y\otimes\mathbb{I}_2-\mathbb{I}_2\otimes\sigma_y)(p_y-\ii
m \omega\sigma_z\otimes\sigma_zy)\\
+&mc^2(\sigma_z\otimes\mathbb{I}_2+\mathbb{I}_2\otimes\sigma_z).
\end{split}
\end{equation}

\section{Energy spectrum and eigenstates}
\label{sectionII}

\begin{figure}[!hbp]

\centering

\begin{overpic}[width=8.5cm]{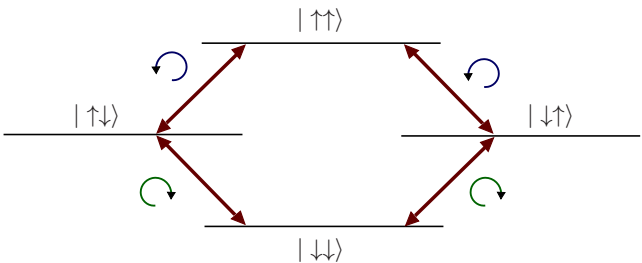}
\put(46,38){{$\ket{\uparrow\uparrow}$}}
\put(46,2){{$\ket{\downarrow\downarrow}$}}
\put(11,23){{$\ket{\uparrow\downarrow}$}}
\put(82,23){{$\ket{\downarrow\uparrow}$}}

\end{overpic}
\caption{Fermionic spin-flip transitions due to the Dirac string
coupling and mediated by the creation-annihilation of chiral
phonons.}\label{niveles}

\end{figure}

In two dimensions, chiral creation-annihilation operators which
carry dual aspects of a left- or right-handed symmetry are defined
as follows
\begin{equation}
\label{circular_operators}
\begin{array}{c}
  a_r:=\frac{1}{\sqrt{2}}(a_x - \ii a_y),\hspace{2ex}a_r^\dagger:=\frac{1}{\sqrt{2}}(a_x^\dagger + \ii a_y^\dagger) , \\
  a_l:=\frac{1}{\sqrt{2}}(a_x + \ii a_y), \hspace{2ex}a_l^\dagger:=\frac{1}{\sqrt{2}}(a_x^\dagger - \ii a_y^\dagger) , \\
\end{array}
\end{equation}
where $ a_x^\dagger,a_x,  a_y^\dagger, a_y$, are the usual
creation-annihilation operators of the harmonic oscillator
$a^{\dagger}_i=\frac{1}{\sqrt{2}}\left(\frac{1}{\tilde{\Delta}}r_i
- \ii \frac{\tilde{\Delta}}{\hbar}p_i\right)$, and
$\tilde{\Delta}=\sqrt{\hbar/m\omega}$ is related to the ground
state width. Using these operators, the relativistic Hamiltonian
in Eq.~\eqref{2D_2body_DO} takes a simpler and amenable form
\begin{equation}
\label{dirac_hamiltonian_matrix}
H_{\text{2D}}=\left[%
\begin{array}{cccc}
  \Delta                & g^* a_l^{\dagger}   & g a_l^{\dagger}     & 0 \\
  g a_l                 & 0                   & 0                   &  g^* a_r  \\
  g^* a_l               & 0                   & 0                   & g a_r  \\
  0                     & g a_r^{\dagger}     & g^* a^{\dagger}_r    & -\Delta \\
\end{array}%
\right],
\end{equation}
where $\Delta:=2mc^2$ stands for the system rest mass,  $g:=\ii
mc^2\sqrt{2\zeta}$ is a coupling parameter, and
$\zeta:=\hbar\omega/mc^2$ controls the strength of the effective
interaction. Considering the two-body spinorial basis
$\{\ket{\uparrow\uparrow},\ket{\uparrow\downarrow},\linebreak
\ket{\downarrow\uparrow},\ket{\downarrow\downarrow}\}$, we can
understand the Dirac string  coupling as a four-level system
depicted in fig.~\ref{niveles}.

\begin{figure}[!hbp]

\centering

\begin{overpic}[width=8.5cm]{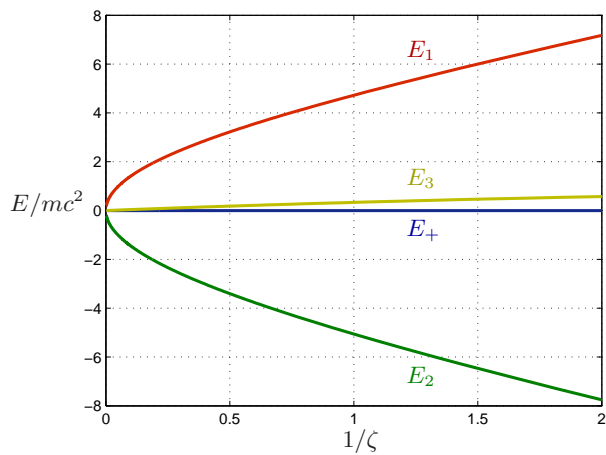}
\put(-2,39){{$E/mc^2$}}
 \put(50,2){{$1/\zeta$}}
\put(60,12){{$\textcolor[rgb]{0.00,0.51,0.00}{E_2}$}}
\put(60,35){{$\textcolor[rgb]{0.00,0.00,0.63}{E_+}$}}
\put(60,43){{$\textcolor[rgb]{0.62,0.62,0.00}{E_3}$}}
\put(60,63){{$\textcolor[rgb]{0.69,0.00,0.00}{E_1}$}}

\end{overpic}
\caption{Dependence  of the two-fermion  energy spectrum with the
dirac string coupling strength. Note that as $1/\zeta\to0$ we
approach the strong coupling regime $\zeta\gg1$, whereas
$1/\zeta\to\infty$ represents a weak attractive coupling
$\zeta\ll1$.} \label{energy_level_figure}
\end{figure}

We now proceed to describe the energy spectrum of the 2-body
interacting relativistic system, in terms of the phonon Fock states
\begin{equation}
\label{chiral_Fock}
\ket{n_r,n_l}:=\frac{1}{\sqrt{n_r!n_l!}}(a_r^{\dagger})^{n_r}(a_l^{\dagger})^{n_l}\ket{\text{vac}},
\end{equation}
where $n_r,n_l=0,1...$ specify the number of right- and
left-handed phonons coupling the two-fermion system. One immediately
observes that the Hilbert space can be divided in a series of
invariant subspaces
$\mathcal{H}=\bigoplus_{n_r,n_l=0}^{\infty}\mathcal{H}_{n_rn_l}$,
where each subspace can be described by
$\mathcal{H}_{n_rn_l}:=\mathcal{H}'_{n_r,n_l}\bigoplus\mathcal{H}''_{n_r,n_l}$.
These subspaces are spanned by
\begin{equation}
\label{hilbert_subspaces}
\begin{split}
\mathcal{H}'_{n_r,n_l}=&\text{span}\{\ket{+}\ket{n_r,n_l}\},\\
\mathcal{H}''_{n_r,n_l}=&\text{span}\{\ket{\uparrow\uparrow}\ket{n_r,n_l+1},\ket{-}\ket{n_r,n_l},\ket{\downarrow\downarrow}\ket{n_r+1,n_l}\},
\end{split}
\end{equation}
where the states
$\ket{-}:=(\ket{\uparrow\downarrow}-\ket{\downarrow\uparrow})/\sqrt{2}$
, and
$\ket{+}:=(\ket{\uparrow\downarrow}+\ket{\downarrow\uparrow})/\sqrt{2}$
are maximally entangled unpolarized Bell states. In particular,
$\mathcal{H}'_{n_r,n_l}$ describes a zero-energy subspace
$E_{+,n_r,n_l}=0$. The
Hamiltonian~\eqref{dirac_hamiltonian_matrix} in the remaining
subspaces $\mathcal{H}''_{n_r,n_l}$ can be expressed as follows
\begin{equation}
\label{dirac_hamiltonian_matrix_3}
H_{\text{2D}}^{n_r n_l}=\Delta\left[%
\begin{array}{ccc}
  1& -\ii\sqrt{\zeta(n_l+1)} & 0 \\
  \ii\sqrt{\zeta(n_l+1)} & 0 & -\ii\sqrt{\zeta(n_r+1)} \\
  0 & \ii\sqrt{\zeta(n_r+1)} & -1\\
\end{array}%
\right],
\end{equation}
where the 2-body interaction couples three different levels and
can be exactly diagonalized. Using Cardano-Vietta solution to
third order polynomials, we obtain the following energies
\begin{equation}
\label{energy_level}
\begin{split}
\frac{E_{1n_rn_l}}{\Delta}:=&\sqrt{\frac{4\left[1+\zeta(n_r+n_l+2)\right]}{3}}\cos\Theta,\\
\frac{E_{2n_rn_l}}{\Delta}:=&\sqrt{\frac{4[1+\zeta(n_r+n_l+2)]}{3}}\cos\left(\Theta+\frac{2\pi}{3}\right),\\
\frac{E_{3n_rn_l}}{\Delta}:=&\sqrt{\frac{4[1+\zeta(n_r+n_l+2)]}{3}}\cos\left(\Theta+\frac{4\pi}{3}\right),\\
\end{split}
\end{equation}
where
\begin{equation}
\Theta:=\frac{1}{3}\text{arccos}\left[\frac{27(n_l-n_r)\zeta}{2[3(1+\zeta(n_r+n_l+2))]^{3/2}}\right].
\end{equation}
These eigenstates are represented for different values of the
coupling strength $\zeta$ in Fig.~\ref{energy_level_figure}, where
the chiral quantum numbers have been set to $n_r=n_l+1$.

In this figure we observe two different regimes:

\noindent \textbf{Weak Coupling regime $\zeta\ll1$:} In this case,
 the low energy properties can be accurately described by a two-level system.
  This feature will
turn out to be crucial for the analogies of the system to a
non-relativistic Cooper pair discussed in
section~\ref{sectionIII}.

\noindent \textbf{Strong Coupling regime $\zeta\gg1$:} In this
case, the four levels become essential in order to describe the
low energy excitations. Consequently, the description becomes more
involved but also gives a richer structure that may show certain
novel properties with respect to non-relativistic Cooper pairs
that are described in section~\ref{sectionIV}.

Once the eigenvalues have been obtained, we may derive the
corresponding eigenstates, which we list below
\begin{equation}
\label{eigenstates}
\begin{split}
\ket{E_{+,n_r,n_l}}:=\ket{+,&n_r,n_l},\\
\ket{E_{j,n_r,n_l}}:=\frac{1}{\Omega_j}[&\alpha_j\ket{-}\ket{n_r,n_l}+\ii\beta_j\ket{\uparrow\uparrow,n_r,n_l+1}\\
+&\ii\delta_j\ket{\downarrow\downarrow,n_r+1,n_l}],
\end{split}
\end{equation}
where we have defined the following parameters
\begin{equation}
\label{norm_constants}
\begin{split}
&\alpha_j:=\Delta^2-E_{jn_rn_l^2}^2,\\
&\beta_j:=\Delta(\Delta+E_{jn_rn_l})\sqrt{\zeta(n_l+1)},\\
&\delta_j:=\Delta(\Delta-E_{jn_rn_l})\sqrt{\zeta(n_r+1)},\\
&\Omega_j:=\sqrt{\alpha_j^2+\beta_j^2+\delta_j^2}.\\
\end{split}
\end{equation}
Here the indexes $j=1,2,3$ correspond to the three different
eigenvalues~\eqref{energy_level}. Let us mention that the total
angular momentum $J_z:= S_z + L_z$ is conserved. Thus, the
eigenstates \eqref{eigenstates} have well-defined angular momenta,
namely, $\hbar(n_r-n_l)$. Finally, we must consider the
consequences of fermion indistinguishability. The Symmetrization
postulate states that a system of identical fermions must be
described in terms of antisymmetrical states, which establishes
the following constraint
\begin{equation}
\label{antisymmetry}
\text{P}_{21}\ket{\Psi(1,2)}=-\ket{\Psi(1,2)},
\end{equation}
where $\text{P}_{21}$ stands for the permutation operator that
swaps the fermion labels $1\leftrightarrow2$. Considering the
eigenstates in Eq.\eqref{eigenstates} under the permutation
operator, we obtain the following expressions
\begin{equation}
\label{permutation_eigenstates}
\begin{split}
\text{P}_{21}\ket{E_{+,n_r,n_l}}&=(-1)^{n_r+n_l}\ket{E_{+,n_r,n_l}},\\
\text{P}_{21}\ket{E_{j\hspace{0.5ex},n_r,n_l}}&=(-1)^{n_r+n_l+1}\ket{E_{j,n_r,n_l}}.
\end{split}
\end{equation}
Since these expressions must satisfy the antisymmetric condition
in Eq.\eqref{antisymmetry}, the number of chiral quanta are
constrained as follows
\begin{equation}
\begin{split}
\ket{E_{+,n_r,n_l}}&\Rightarrow n_r+n_l=2k+1\hspace{2ex}:
k=0,1,2...\\
\ket{E_{j\hspace{0.5ex},n_r,n_l}}&\Rightarrow
n_r+n_l=2k\hspace{6ex}:
k=0,1,2...\\
\end{split}
\end{equation}
Due to the indistinguishability of the relativistic fermions, the
eigenstates $\ket{E_{+,n_r,n_l}}$ must contain an odd number of
chiral quanta, whereas $\ket{E_{j,n_r,n_l}}$ are restricted to
even number of chiral quanta.

We have thus derived a complete solution of the relativistic Dirac
equation for two bodies interacting via a Dirac string coupling.
Therefore, this 2-fermion system  belongs to the small class of
exactly solvable few-body relativistic systems. In
sections~\ref{sectionIII} and~\ref{sectionIV} we show that this
relativistic interaction does indeed lead to the formation of
bound pairs, both in the weak and strong coupling regimes.
Furthermore, we present a detailed study of the similar properties
that the relativistic bound pair shares with the well-known
non-relativistic Cooper pair. As we will see, there are profound
analogies in the weak coupling regime, whereas novel properties
are found in the strong coupling limit.

\section{Weak Coupling Regime}
\label{sectionIII}

 The standard description of Cooper pairs in superconducting solids
 is usually performed in a weak coupling regime, where a slightly
 phonon-mediated attractive interaction  binds electron which
 lay close to the Fermi surface. We shall consider that the
 two-body Hamiltonian in Eq.~\eqref{2D_2body_DO} effectively
 describes the gluing mechanism above the Fermi sea, and therefore
 a weak coupling regime is obtained when $\zeta\ll1$.

\begin{figure}[!hbp]

\centering

\begin{overpic}[width=8.5cm]{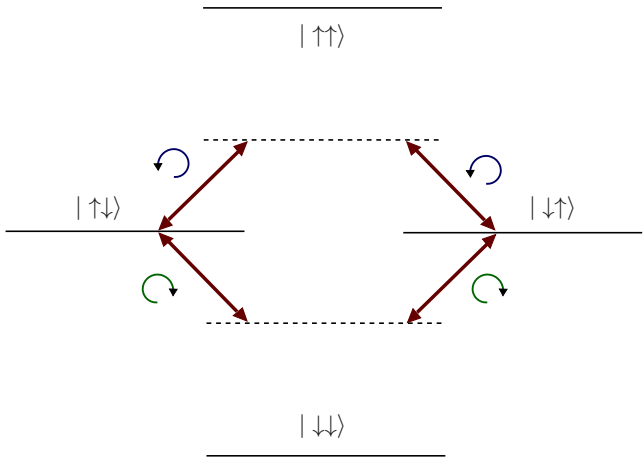}
\put(46,65){{$\ket{\uparrow\uparrow}$}}
\put(46,4){{$\ket{\downarrow\downarrow}$}}
\put(11,38){{$\ket{\uparrow\downarrow}$}}
\put(82,38){{$\ket{\downarrow\uparrow}$}}

\end{overpic}
 \caption{Fermionic spin-flip transitions in the weak coupling regime. The low energy sector is described by an
 effective two-level system, where  spin-flips occur along two different channels that include
 virtual two-phonon transitions and spin-polarized states become decoupled.}\label{2_niveles}
\end{figure}

In this weak coupling regime, we have seen in
Fig.~\ref{energy_level_figure} that the low-lying excitations can
be entirely described by a two-level system. This situation is
schematically described fig.~\ref{2_niveles}, where we see how
spin-polarized levels become decoupled from those responsible of the
low-energy properties. In this situation, we can obtain an effective
Hamiltonian for the low energy sector, by adiabatic elimination.

Let us consider an arbitrary state
$\ket{\Psi(t)}\in\mathcal{H}_{n_rn_l}$
\begin{equation}
\begin{split}
\ket{\psi(t)}=&c_{\uparrow\uparrow}(t)\ket{\uparrow\uparrow,n_r,n_l+1}+c_{\uparrow\downarrow}(t)\ket{\uparrow\downarrow,n_r,n_l}\\
&+c_{\downarrow\uparrow}(t)\ket{\downarrow\uparrow,n_r,n_l}+c_{\downarrow\downarrow}(t)\ket{\downarrow\downarrow,n_r+1,n_l},
\end{split}
\end{equation}
whose dynamical evolution, described by the Dirac
Hamiltonian~\eqref{dirac_hamiltonian_matrix}, can be represented
as
\begin{equation}
\label{Hamiltonian_dynamics}
\ii\hbar\frac{\dd}{\dd t}\left[%
\begin{array}{c}
  c_{\uparrow\uparrow}(t) \\
  c_{\uparrow\downarrow}(t) \\
  c_{\downarrow\uparrow}(t) \\
  c_{\downarrow\downarrow}(t) \\
\end{array}%
\right]=\left[%
\begin{array}{cccc}
  \Delta                & g^* a_l^{\dagger}   & g a_l^{\dagger}     & 0 \\
  g a_l                 & 0                   & 0                   &  g^* a_r  \\
  g^* a_l               & 0                   & 0                   & g a_r  \\
  0                     & g a_r^{\dagger}     & g^* a^{\dagger}_r    & -\Delta \\
\end{array}%
\right]\left[%
\begin{array}{c}
  c_{\uparrow\uparrow}(t) \\
  c_{\uparrow\downarrow}(t) \\
  c_{\downarrow\uparrow}(t) \\
  c_{\downarrow\downarrow}(t) \\
\end{array}
\right].
\end{equation}
In the weak coupling limit, the transitions to the spin-polarized
$\{\ket{\uparrow\uparrow,n_r,n_l+1},\ket{\downarrow\downarrow,n_r+1,n_l}\}$
upper and lower levels can be considered negligible. Therefore,
the level population does not evolve under the action of the
two-body interaction $\frac{\dd c_{\uparrow\uparrow}}{dt}=\frac{d
c_{\downarrow\downarrow}}{\dd t}=0$, and we can adiabatically
eliminate these two levels. The latter conditions substituted in
Eq.~\eqref{Hamiltonian_dynamics}, give rise to the following
relations
\begin{equation}
\begin{split}
c_{\uparrow\uparrow}&=\ii\sqrt{\frac{\zeta(n_l+1)}{2}}(c_{\uparrow\downarrow}-c_{\downarrow\uparrow}),\\
c_{\downarrow\downarrow}&=\ii\sqrt{\frac{\zeta(n_r+1)}{2}}(c_{\uparrow\downarrow}-c_{\downarrow\uparrow}),
\end{split}
\end{equation}
and an effective two-level dynamics
\begin{equation}
\label{two_level_Hamiltonian_dynamics}
\ii\hbar\frac{\dd}{\dd t}\left[%
\begin{array}{c}
  c_{\uparrow\downarrow}(t) \\
  c_{\downarrow\uparrow}(t) \\
\end{array}%
\right]=\frac{\Delta\zeta}{2}(n_r-n_l)\left[%
\begin{array}{cc}
  \hspace{1.5ex}1                & -1   \\
  -1                 & \hspace{1.5ex}1                    \\
\end{array}%
\right]\left[%
\begin{array}{c}
  c_{\uparrow\downarrow}(t) \\
  c_{\downarrow\uparrow}(t) \\
\end{array}%
\right].
\end{equation}
In this sense, we can integrate out the high-frequency modes by
projecting onto the effective spin-unpolarized invariant subspace
spanned by
$\mathcal{H}^{\text{eff}}_{n_rn_l}:=\text{span}\{\ket{\uparrow\downarrow,n_r,n_l},\ket{\downarrow\uparrow,n_r,n_l}\}$,
by means of an orthogonal projector
\begin{equation}
\mathcal{P}^{\text{eff}}_{n_rn_l}:=\ket{\uparrow\downarrow,n_r,n_l}\bra{\uparrow\downarrow,n_r,n_l}+\ket{\downarrow\uparrow,n_r,n_l}\bra{\downarrow\uparrow,n_r,n_l}
\end{equation}
The $z$-component of the orbital angular momentum operator
$L_z=\hbar(a_r^{\dagger}a_r-a_l^{\dagger}a_l)$ constrained to this
invariant subspace becomes $
\mathcal{P}^{\text{eff}}_{n_rn_l}L_z\mathcal{P}^{\text{eff}}_{n_rn_l}=\hbar(n_r-n_l)\mathbb{I}_2,
$ which allows us to
 rewrite Eq.~\eqref{two_level_Hamiltonian_dynamics} as an
 effective two-level Hamiltonian
 \begin{equation}
 \label{raman_DO}
 H_{\text{eff}}:=\omega L_z \left[%
\begin{array}{cc}
  \hspace{1.5ex}1                & -1   \\
  -1                 & \hspace{1.5ex} 1                    \\
\end{array}%
\right]=\hbar\omega(a_r^{\dagger}a_r-a_l^{\dagger}a_l)\left[%
\begin{array}{cc}
  \hspace{1.5ex}1                & -1   \\
  -1                 &\hspace{1.5ex} 1                    \\
\end{array}%
\right].
 \end{equation}
 This effective interaction in the weak coupling regime is represented in
Fig.~\ref{2_niveles}, where the allowed transitions can take on
two different channels via the consecutive creation-annihilation
of right- or left-handed phonons.
 This process can be understood
as an instance of a superexchange coupling between the spins
$\ket{\uparrow\downarrow}\longleftrightarrow\ket{\downarrow\uparrow}$
driven by a second order two-phonon process where a chiral phonon
is virtually created and then annihilated. There exist two
different exchange paths, as seen in fig.~\ref{2_niveles},
depending on the left- or right-handed chiralities of the virtual
phonons involved in the process.

 This effective
Hamiltonian~\eqref{raman_DO} can be exactly diagonalized yielding
the eigenvalues
\begin{equation}
E_{+n_rn_l}^{\text{eff}}:=0,\hspace{2ex}E_{-n_rn_l}^{\text{eff}}:=2\hbar\omega(n_r-n_l),
\end{equation}
with the following associated eigenstates
\begin{equation}
\label{NR_eigenstates}
\begin{split}
\ket{E_{+n_rn_l}^{\text{eff}}}:=\ket{+,n_r,n_l}&\Rightarrow
n_r+n_l=2k+1\hspace{2ex}:
k=0,1,2...\\
\ket{E_{-n_rn_l}^{\text{eff}}}:=\ket{-,n_r,n_l}&\Rightarrow
n_r+n_l=2k\hspace{6ex}:
k=0,1,2...\\
\end{split}
\end{equation}
where the anti-symmetric character of the fermionic states has
already been considered. Therefore, the low-lying solution in the
weak coupling regime can be described by the maximally entangled
Bell states in the spin degree of freedom, and rotational Fock
states in the orbital degree of freedom.

Furthermore, these states describe a bound fermion pair. In order
to show that such binding occurs, we must show that the
inter-particle distance only attains finite values. Let us
introduce the square-distance operator $ \Gamma:=x^2+y^2, $ where
$x:=(x_1-x_2)/\sqrt{2}$ and $y:=(y_1-y_2)/\sqrt{2}$ denote the
space coordinate operators for the relative fermionic distance.
The expectation values  in the weak-coupling
eigenstates~\eqref{NR_eigenstates} are
\begin{equation}
\label{two-level-confinement}
 \langle\Gamma\rangle_{\pm}=\frac{\tilde{\Delta}^2}{\sqrt{2}}(1+n_r+n_l),
\end{equation}
which is always finite. We observe the crucial property that this
system shares with a non-relativistic Cooper pair, namely, the
pair of relativistic fermions are bounded in pairs even for a weak
attraction $\zeta\ll 1$.

Another fundamental property that occurs in standard Cooper pairs
is the presence of an energy gap between the paired energy level
and the Fermi surface. This energy gap is responsible of the
stability of Cooper pairs with respect to free fermion pairs and
is proportional to the lattice Debye frequency $\Delta
E\sim\hbar\omega_D$. In the relativistic regime, we observe that
the energy gap with respect to the displaced Fermi surface ( i.e.
$\epsilon_F'=0$ ) is
\begin{equation}
\begin{split}
&\Delta E_{+n_rn_l}=0, \\
&\Delta E_{-n_rn_l}=2\hbar\omega(n_r-n_l),
\end{split}
 \end{equation}
and therefore the only stable pair (i.e. $\Delta E<0$) is that
described by the spin-singlet state when $n_l\geq n_r$.

In this sense we obtain a spin-singlet bound pair which clearly
resembles the situation in standard Cooper pairs where the
fermions are also in the singlet state. Furthermore, we can
observe from this discussion that the relativistic gap is
proportional to the Dirac string frequency $\Delta
E\sim\hbar\omega$, which plays the role of the usual Debye
frequency in superconducting materials.

Finally, to take this comparison further, we should study the
properties of the stable pair eigenstates in
Eq.~\eqref{NR_eigenstates} and compare them to the
non-relativistic Cooper pair features.

\textbf{Spin degrees of freedom:} In BCS theory, Cooper pairs
display a singlet state in the spin degree of freedom. We observe
in Eq.~\eqref{NR_eigenstates} that the stable bound fermionic pair
state has also a spin-singlet component.

\begin{figure}[!hbp]

\centering

\begin{overpic}[width=8.5cm]{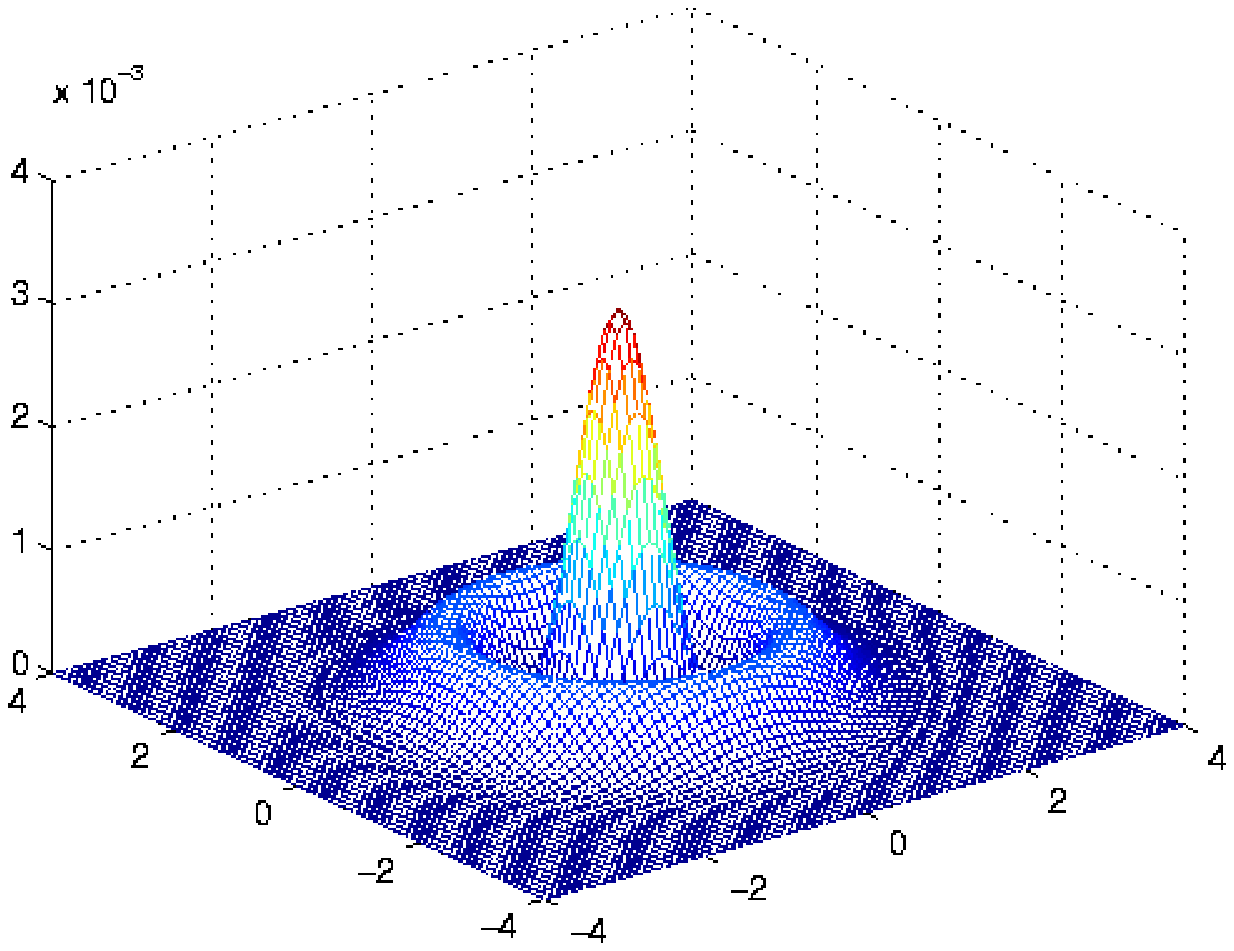}
\put(18,12){{$y/\tilde{\Delta}$}} \put(75,9){{$x/\tilde{\Delta}$}}
\put(-2,41){{$\rho^{\text{eff}}_{-11}$}}

\end{overpic}

\begin{overpic}[width=8.5cm]{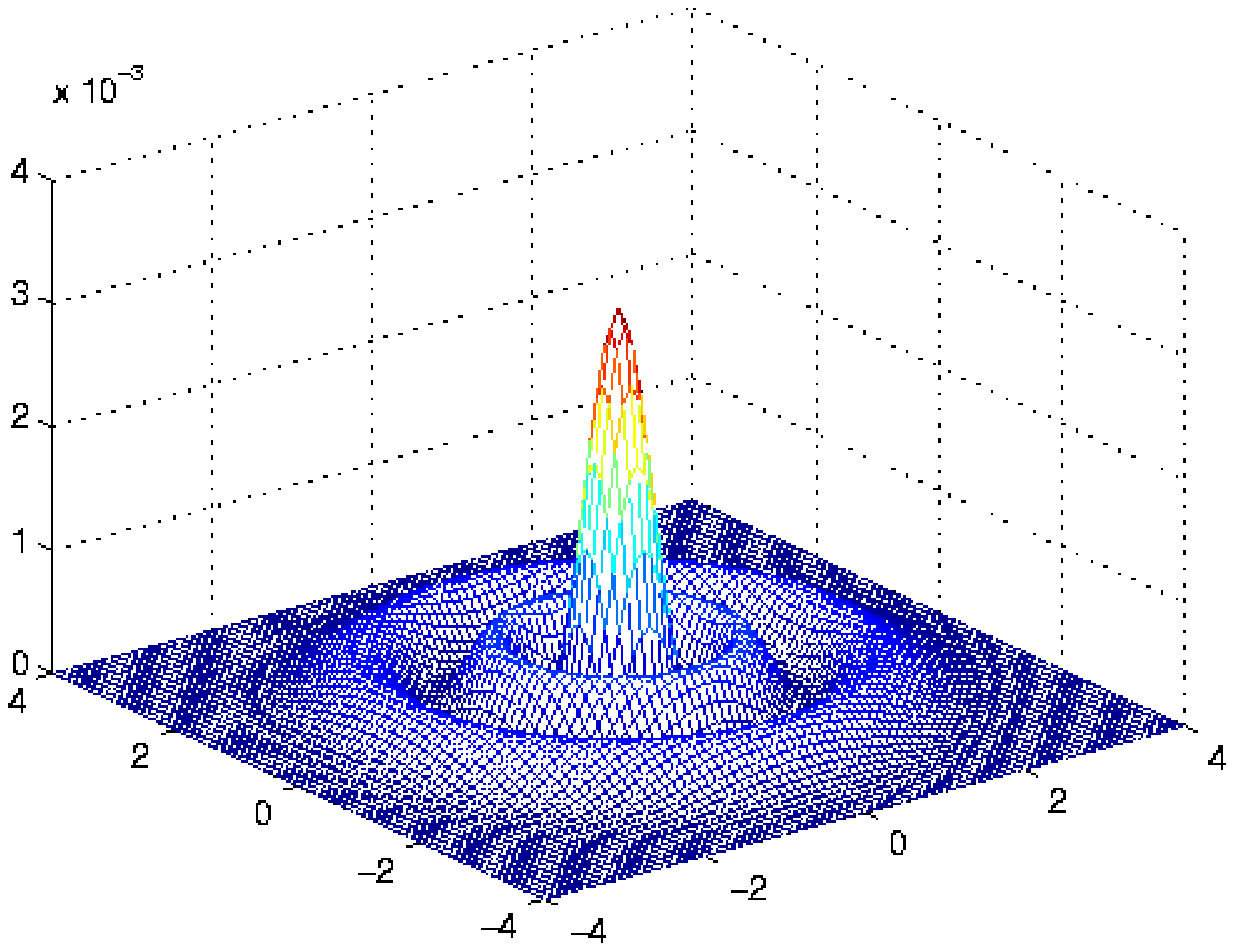}
\put(18,12){{$y/\tilde{\Delta}$}} \put(75,9){{$x/\tilde{\Delta}$}}
\put(-2,41){{$\rho^{\text{eff}}_{-22}$}}

\end{overpic}

\caption{Spatial probability density profiles for weak coupling
stable pairs
$\rho^{\text{eff}}_{-n_rn_l}(r):=\text{Tr}_{\text{spin}}(\langle\textbf{r}\ket{E_{-n_rn_l}^{\text{eff}}}\bra{E_{-n_rn_l}^{\text{eff}}}\textbf{r}\rangle)$
with $n_l>n_r$. Top figure corresponds to $n_r=1,n_l=1$
probability density $\rho^{\text{eff}}_{-11}(r)$, with
$r=|\textbf{r}_1-\textbf{r}_2|/\sqrt{2}.$ Bottom figure represents
$\rho^{\text{eff}}_{-22}(r)$.} \label{fock_states_1_profile}
\end{figure}

\textbf{Orbital degrees of freedom:} In BCS theory, Cooper pairs
display a spherically symmetrical wave function with an onion-like
layered structure. We directly observe from
fig.~\ref{fock_states_1_profile} that relativistic bound pair
probability distribution $\rho^{\text{eff}}_{-n_rn_l}(r)$ display
a similar spherically symmetric onion-like structure.

In this section we have discussed a relativistic pairing mechanism
in a weak coupling regime. We have discussed in detail several
analogies with a non-relativistic Cooper pair that naturally arise
in this weak coupling limit. Remarkably, we obtain binding
regardless of the interaction strength, which is a fundamental
property of BCS systems. Additionally, we have shown how the
relativistic energy gap scales with the string frequency in the
same manner as the Cooper pair gap scales with the phonon Debye
frequency. In this regard, we may conclude that the string
interaction plays the role of the lattice phonons that mediate the
effective attraction between fermions in  the BCS theory.
Furthermore, we have also compared the nature of the relativistic
pair eigenstates with the Cooper pair wave functions. We have seen
that the relativistic bound pair is also described by a
spin-singlet state and a spherically symmetric onion-like state in
the orbital degrees of freedom. All these similarities allow us to
state that this fermionic pairing mechanism can be interpreted as
a relativistic Cooper pair, since we recover most of the usual BCS
properties in the weak coupling regime. Nonetheless, this
Relativistic Cooper pair can also be studied in the strong
coupling regime, where novel properties
 with respect to the usual Cooper pair in BCS theory arise. As we
 describe below, when the Dirac string interaction becomes strong
 enough, phonons contribute dynamically to the gluing mechanism.

\section{Strong Coupling Regime}
\label{sectionIV}

\begin{figure}[!hbp]

\centering

\begin{overpic}[width=9cm]{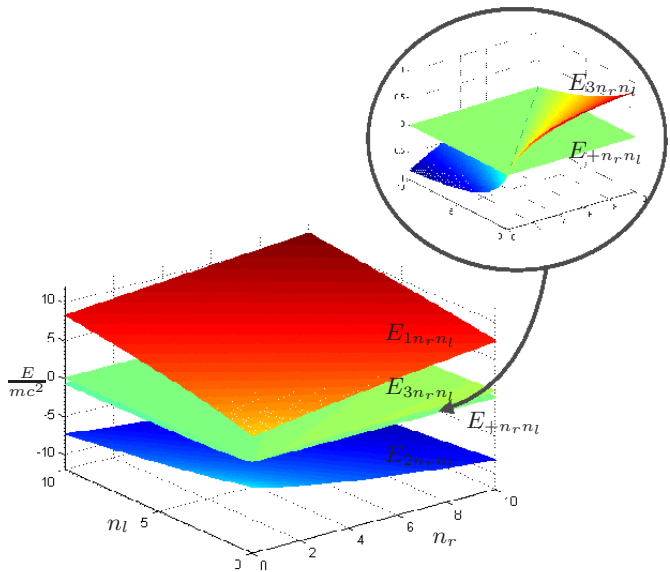}
\put(2,31){{$\frac{E}{mc^2}$}} \put(17,10){{$n_l$}}
\put(65,8){{$n_r$}} \put(58,20){{$E_{2n_rn_l}$}}
\put(58,38){{$E_{1n_rn_l}$}} \put(58,30){{$E_{3n_rn_l}$}}
\put(70,25){{$E_{+n_rn_l}$}} \put(85,65){{$E_{+n_rn_l}$}}
\put(85,75){{$E_{3n_rn_l}$}}

\end{overpic}
 \caption{Energy levels of the two-body Dirac oscillator in the strong coupling regime
 $\zeta=5$ as a function of the different phonon number.(inset) Detail of the energies corresponding to levels
 $E_{+n_rn_l}, E_{3n_rn_l}$}
\label{energy_levels_strong_coupling}
\end{figure}

In this section we study the pairing properties of the two-body
relativistic system in the strong coupling regime $\zeta\gg1$. In
this limit we must consider the complete four-level structure of
the system ( see fig.~\ref{niveles} ), and the energy spectrum
becomes clearly more involved in Eq.~\eqref{energy_level} (see
fig.~\ref{energy_levels_strong_coupling}).

In fig.~\ref{energy_levels_strong_coupling} we have represented
the different energies for an interaction strength $\zeta=5$ which
lays in the strong coupling regime. We clearly see how two levels
$E_{2,n_r,n_l},E_{3,n_r,n_l}$ become stable pairs with a certain
gap $\Delta E_{2,n_r,n_l}<\Delta E_{3,n_r,n_l}<0$. Therefore the
strong coupling gives raise to a couple of stable bound fermionic
states, namely,
\begin{equation}
\label{cooper_pair_strong_coupling}
\begin{split}
\ket{E_{2,n_r,n_l}}:=\frac{1}{\Omega_2}[&\alpha_2\ket{-}\ket{n_r,n_l}+\ii\beta_2\ket{\uparrow\uparrow,n_r,n_l+1}+\\
+&\ii\delta_2\ket{\downarrow\downarrow,n_r+1,n_l}];\\
\ket{E_{3,n_r,n_l}}:=\frac{1}{\Omega_3}[&\alpha_3\ket{-}\ket{n_r,n_l}+\ii\beta_3\ket{\uparrow\uparrow,n_r,n_l+1}+\\
+&\ii\delta_3\ket{\downarrow\downarrow,n_r+1,n_l}].
\end{split}
\end{equation}

\begin{figure}[!hbp]

\centering
\begin{overpic}[width=8.5cm]{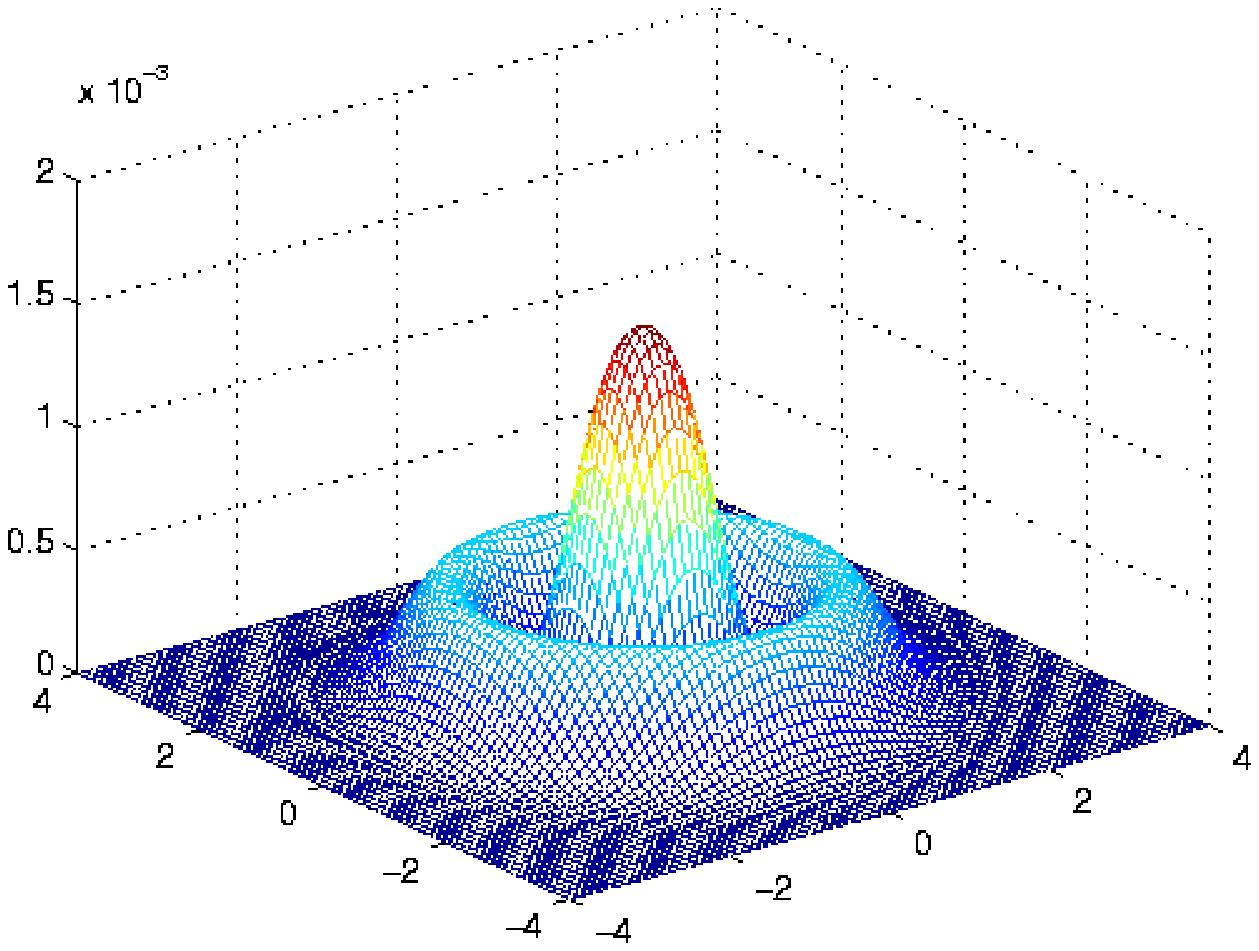}
\put(18,12){{$y/\tilde{\Delta}$}} \put(75,9){{$x/\tilde{\Delta}$}}
\put(-2,41){{$\rho_{211}$}}

\end{overpic}

\begin{overpic}[width=8.5cm]{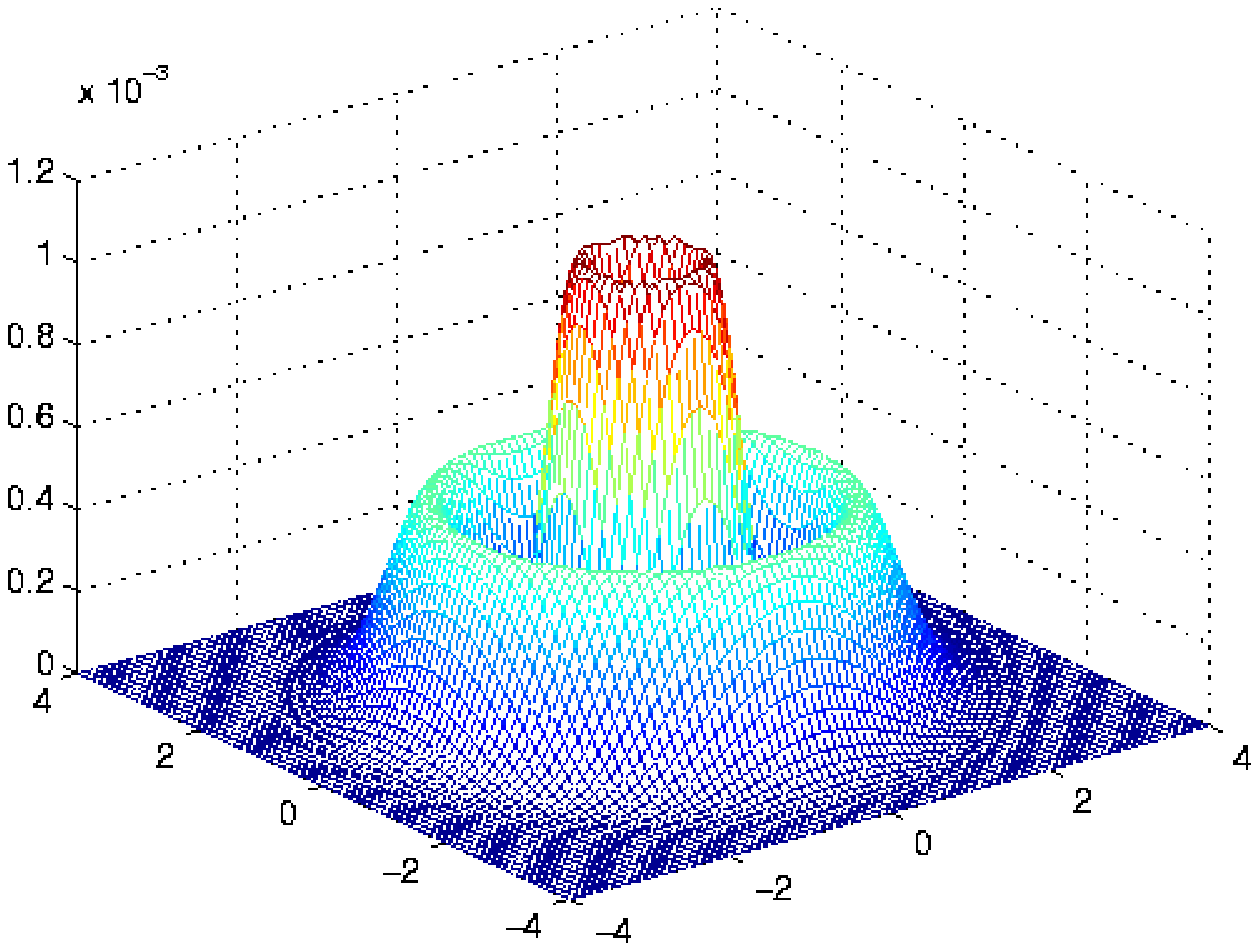}
\put(18,12){{$y/\tilde{\Delta}$}} \put(75,9){{$x/\tilde{\Delta}$}}
\put(-2,41){{$\rho_{311}$}}

\end{overpic}

\caption{Spatial probability density profiles for strong coupling
$\zeta=5$ stable pairs
$\rho_{jn_rn_l}(r):=\text{Tr}_{\text{spin}}(\langle\textbf{r}\ket{E_{jn_rn_l}}\bra{E_{jn_rn_l}}\textbf{r}\rangle)$
with  $n_r=1,n_l=1$. Top figure corresponds to the probability
density of the stable pair $\rho_{211}(r)$, with
$r=|\textbf{r}_1-\textbf{r}_2|/\sqrt{2}.$ Bottom figure represents
 the probability density of the stable
pair $\rho_{311}(r)$.} \label{fock_states_2_profile}
\end{figure}

 The spatial probability distribution $\rho_{jn_rn_l}(r)$ of these stable
 fermionic pairs has been represented in
 fig.~\ref{fock_states_2_profile} in the case of $n_r=n_l=1$. We
 can clearly observe that the density profile preserves the
 spherically symmetric onion-like structure. Nonetheless,
 noteworthy differences arise with respect to the weak coupling
 regime ( compare to the top fig.~\ref{fock_states_1_profile}).

Furthermore, these two stable states form a fermionic bound pair
since the inter-particle distance is finite
\begin{equation} \label{binding_distance_strong_coupling}
\begin{split}
\langle\Gamma\rangle_{2\hspace{0.5ex}}&=\tilde{\Delta}^2\left[(1+n_r+n_l)\alpha_2^2+(\beta_2^2+\delta_2^2)(2+n_r+n_l)\right],\\
\langle\Gamma\rangle_{3\hspace{0.5ex}}&=\tilde{\Delta}^2\left[(1+n_r+n_l)\alpha_3^2+(\beta_3^2+\delta_3^2)(2+n_r+n_l)\right].
\end{split}
\end{equation}
We may conclude that the Dirac string pairing mechanism  leads to
bound pairs in the strong coupling regime, which display
substantial differences with respect to the weakly coupled bound
states in Eq.~\eqref{NR_eigenstates}. It follows from
Eqs.~\eqref{cooper_pair_strong_coupling} that the bound pairs are
not in a singlet state but rather in a linear superposition of
different spin singlet and triplet states entangled with different
orbital Fock states. In this regard, the relativistic pairing
mechanism does not induce an anti-ferromagnetic ordering any
longer, and certain spin-polarization may arise depending on the
value of the coupling strength $\zeta$.

It is also instructive to compare the orbital degrees of freedom
of bound pairs in the weak and strong coupling limits. The weakly
coupled states in Eq.~\eqref{NR_eigenstates} are in orbital Fock
states, which represent a certain number of vibrational phonons
which are frozen in this limit. On the other hand, strongly
coupled states in Eqs.~\eqref{cooper_pair_strong_coupling} cannot
be described by a single Fock state, and therefore the vibrational
phonons acquire a dynamical behavior
$\ket{n_r+1,n_l}\leftrightarrows\ket{n_r,n_l}\leftrightarrows\ket{n_r,n_l+1}$,
which is a clear sign of strong coupling in superconductors. We
may conclude that the Dirac string phonons, responsible of the
gluing mechanism, become unfrozen as the coupling becomes stronger
and contribute to the effective attraction in a dynamic
phenomenon. This is reminiscent of a $(s,p)$-wave symmetry of a SC
order parameter. Similar types of superconducting states appear in
some quantum liquids like superfluid ${\rm He}^3$: the so-called
A- and B-phases exhibit different patterns of spin-orbit symmetry
breaking \cite{helium-3,leggett}. Layered materials like the
ruthenates also exhibit  unusual symmetry properties like triplet
superconductivity \cite{maeno,rice_sigrist,baskaran}.

We also observe that the strong pairing mechanism leads to a
couple of possible stable bound
pairs~\eqref{cooper_pair_strong_coupling}, whereas the weak
coupling only produces one stable bound pair. Furthermore, the
energy gap displayed by the bound pairs also depends on the
strength of the coupling. In the weak coupling regime, we have
already seen that the energy gap scales as $\Delta
E_-\sim\hbar\omega$, whereas the scaling in the strongly
interaction limit does not present such a simple scaling (see
fig.~\ref{energy_levels_strong_coupling}).

\section{Conclusions}

We have studied the relativistic pairing mechanism of the two-body
Dirac oscillator in two dimensions, where a Dirac string coupling
leads to fermionic bound pairs. We have described two different
regimes where binding occurs regardless of the interaction
strength.

In a weak coupling regime, the fermionic pair bears a strong
resemblance to the usual Cooper pair in BCS theory. We remarkably
found a similar scaling of the energy gap, which allows us to
identify the Dirac string frequency $\omega$ with the lattice
phonon Debye frequency $\omega_{D}$ in superconducting materials.
Additionally, we found that the relativistic bound pair
eigenstates are also in a spin singlet state, and present a
spherically symmetric onion-like structure in the probability
distribution. All these remarkable analogies suggest to interpret
this two-body Dirac oscillator as an instance of a relativistic
Cooper pair. Nevertheless, there may also be other types of
relativistic binding mechanisms yielding also the formation of
Cooper pairs.

On the other hand, a strong interaction leads to remarkable
differences with respect to BCS Cooper pairs. In this case, more
than one bound pair can be built, which in any case is not in a
singlet state but rather in a linear superposition of singlet and
triplet states. Furthermore, the gluing phonons become unfrozen as
the coupling strength is raised and dynamically contribute to the
pairing mechanism.

\noindent {\em Acknowledgements} We acknowledge financial support
from a FPU grant of the MEC (A.B.), DGS grant  under contract
FIS2006-04885, CAM-UCM grant under ref. 910758. (A.B., M.A.M.D,),
and the ESF Science Programme INSTANS 2005-2010 (M.A.M.D.).



\begin{references}




\bibitem{BCS}
J. Bardeen, L. N. Cooper, and J. R. Schrieffer, Phys. Rev.
\textbf{108}, 1175 (1957).

\bibitem{frohlich}
H. Fr\"{o}lich, Phys. Rev. \textbf{79}, 845 (1950).

\bibitem{Cooper}
L. N. Cooper, Phys. Rev. \textbf{104}, 1189 (1956).

\bibitem{greiner_book}
W. Greiner, \emph{``Relativistic Quantum Mechanics. Wave
Equations.''}, (Springer, Berlin, 2000).



\bibitem{imc67}
D. Ito, K. Mori, and E. Carrieri, N. Cimento  {\bf 51 A}, 1119,
(1967).

\bibitem{moshinsky}
M. Moshinsky, and  A. Szcepaniak, J. Phys. A \textbf{22}, L817,
(1989).

\bibitem{barut}
A. O. Barut, and  S. Komy, Fortsch. Phys. \textbf{33}, 6 (1985).

\bibitem{moshinsky_2Body}
M. Moshinsky, G. Loyola, and C. Villegas, J. Math. Phys.
\textbf{32}, 373 (1990).


\bibitem{moshinsky_pp+}
M. Moshinsky, and G. Loyola, Found. Phys. \textbf{23}, 197 (1993).

\bibitem{moshinsky_meson}
A. Gonz\'alez, G. Loyola, and  M. Moshinsky, Rev. Mex. Fis.
\textbf{40}, 12 (1994).

\bibitem{migdal}
A.B. Migdal,
Zh. Eksp. Teor. Fiz., {\bf 34},  1438 (1958);
(Sov. Phys. JETP, {\bf 7},  996 (1958)).

\bibitem{eliashberg}
G.M. Eliashberg,
Zh. Eksp. Teor. Fiz., {\bf 38},  966 (1960);
{\bf 39},  1437 (1960);
(Sov. Phys. JETP, {\bf 11},  696 (1960); {\bf 12}, 1000 (1960)).

\bibitem{carbotte}
J.P. Carbotte,
Rev. Mod. Phys. {\bf 62}, 1027 (1990).

\bibitem{bednorz_muller}
J.G. Bednorz, and K. A. M\"{u}ller,  Z. Phys. B {\bf 64}, 189
(1986).

\bibitem{helium-3}
D.D. Osheroff, R.C. Richardson, and D.M. Lee, Phys. Rev. Lett.
{\bf 28}, 885 (1972).

\bibitem{leggett}
A.J. Leggett, Rev. Mod. Phys. {\bf 47}, 331 (1975).

\bibitem{maeno}
Y. Maeno et al., Nature {\bf 372}, 532 (1994).

\bibitem{rice_sigrist}
T. M. Rice, and M. Sigrist, J. Phys. Cond. Matter {\bf 7}, L643
(1995).

\bibitem{baskaran}
G. Baskaran, Physica B 223-224, 490 (1996).





\end{references}
\end{document}